\documentclass[aps, prx, amsmath,amssymb, reprint, superscriptaddress]{revtex4-2}

\usepackage{graphicx}
\usepackage{bm}
\usepackage{tabularx}
\usepackage{graphicx}
\usepackage{siunitx}
\usepackage{booktabs}
\usepackage{xcolor}
\usepackage[colorlinks=true,citecolor=blue]{hyperref}
\usepackage{natbib}
\usepackage{tablefootnote}

\begin{document}
	
	\title{2D Janus Niobium Oxydihalide NbO$XY$: \\Multifunctional High-Mobility Piezoelectric Semiconductor for Electronics, Photonics\\and Sustainable Energy Applications}
	
	\author{Tong Su}
    \affiliation{Sauvage Laboratory for Smart Materials, School of Materials Science and Engineering, Harbin Institute of Technology, Shenzhen 518055, China}
    \affiliation{Shenzhen Key Laboratory of Flexible Printed Electronics Technology, Harbin Institute of Technology, Shenzhen 518055, China}	\affiliation{Science, Mathematics and Technology, Singapore University of Technology and Design, Singapore 487372}

    \author{Ching Hua Lee}
    \affiliation{Department of Physics, National University of Singapore, Singapore 117542}
    
    \author{San-Dong Guo}
    \affiliation{School of Electronic Engineering, Xi'an University of Posts and Telecommunications, Xi'an 710121, China}
    
    \author{Guangzhao Wang}
    \affiliation{Key Laboratory of Extraordinary Bond Engineering and Advanced Materials Technology of Chongqing, School of Electronic Information Engineering, Yangtze Normal University, Chongqing 408100, China}
    
    \author{Wee-Liat Ong}
    \affiliation{ZJU-UIUC Institute, College of Energy Engineering, Zhejiang University, Jiaxing, Haining, Zhejiang, 314400, China}
    \affiliation{State Key Laboratory of Clean Energy Utilization, Zhejiang University, Hangzhou, Zhejiang, 310027, China}

    \author{Liemao Cao}
    \affiliation{College of Physics and Electronic Engineering, Hengyang Normal University, Hengyang 421002, China}

    \author{Weiwei Zhao}
    \email{Corresponding Author. wzhao@hit.edu.cn}
    \affiliation{Sauvage Laboratory for Smart Materials, School of Materials Science and Engineering, Harbin Institute of Technology, Shenzhen 518055, China}
    \affiliation{Shenzhen Key Laboratory of Flexible Printed Electronics Technology, Harbin Institute of Technology, Shenzhen 518055, China}
    
    \author{Shengyuan A. Yang}
    \affiliation{Science, Mathematics and Technology, Singapore University of Technology and Design, Singapore 487372}	
    
	\author{Yee Sin Ang}
	\email{Corresponding Author. yeesin\_ang@sutd.edu.sg}
	\affiliation{Science, Mathematics and Technology, Singapore University of Technology and Design, Singapore 487372}	

\begin{abstract}

Two-dimensional (2D) niobium oxydihalide NbOI$_2$ has been recently demonstrated as an excellent in-plane piezoelectric and nonlinear optical material. Here we show that Janus niobium oxydihalide, NbO$XY$ (X, Y = Cl, Br, I and X$\neq$Y), is a multifunctional anisotropic semiconductor family with exceptional piezoelectric, electronic, photocatalytic and optical properties. NbO$XY$ are stable and flexible monolayers with band gap around the visible light regime of $\sim 1.9$ eV. The anisotropic carrier mobility of NbO$XY$ lies in the range of $10^3 \sim 10^4$ cm$^2$V$^{-1}$s$^{-1}$, which represents some of the highest among 2D semiconductors of bandgap $\gtrsim 2$ eV. Inversion symmetry breaking in Janus NbO$XY$ generates sizable out-of-plane $d_{31}$ piezoelectric response while still retaining a strong in-plane piezoelectricity. Remarkably, NbO$XY$ exhibits an additional out-of-plane piezoelectric response, $d_{32}$ as large as 0.55 pm/V. G$_0$W$_0$-BSE calculation further reveals the strong linear optical dichroism of NbO$XY$ in the visible-to-ultraviolet regime. The optical absorption peaks of $14\sim18$ \% in the deep UV regime ($5\sim6$ eV) outperform the vast majority of other 2D materials. The high carrier mobility, strong optical absorption, sizable built-in electric field and band alignment compatible with overall water splitting further suggest the strengths of NbO$XY$ in solar-to-hydrogen conversion. We further propose a directional stress sensing device to demonstrate how the out-of-plane piezoelectricity can be harnessed for functional device applications. Our findings unveil NbO$XY$ as an exceptional multifunctional 2D semiconductor for flexible electronics, optoelectronics, UV photonics, piezoelectronics and sustainable energy applications.

\end{abstract}

\maketitle

\section{Introduction}
Piezoelectricity is a phenomenon in which electrical (mechanical) signals are generated in a material in response to an external mechanical (electrical) stimuli.
Piezoelectric materials, including crystals \cite{cui2018two}, polymer \cite{shehzad2021flexible}, bi-molecules \cite{bera2021molecular} and 2D materials \cite{zhang2021piezotronics}, play a critical role in electromechanical and mechanoelectrical device technology, such as sensors and actuators \cite{janshoff2000piezoelectric,kingon2005lead,wang2006piezoelectric,masmanidis2007multifunctional,wu2016piezotronics,khan2016piezoelectric}. 
Alongside with solid-state energy conversion strategies, such as solar cell \cite{solar}, thermoelectricity \cite{super_TE} and triboelectricity \cite{FAN2012328}, piezoelectricity represents another promising contender for salvaging electrical energy from mechanical motion \cite{wu2014piezoelectricity}. 
Materials with simultaneous presence of excellent piezoelectric, electrical, mechanical and optical properties are particularly much sought-after due to their enormous technological usefulness for developing \emph{multifunctional} devices that synergize piezoelectricity with other functionalities. 

\begin{figure*}[t]
    \centering
    \includegraphics[scale=0.5058]{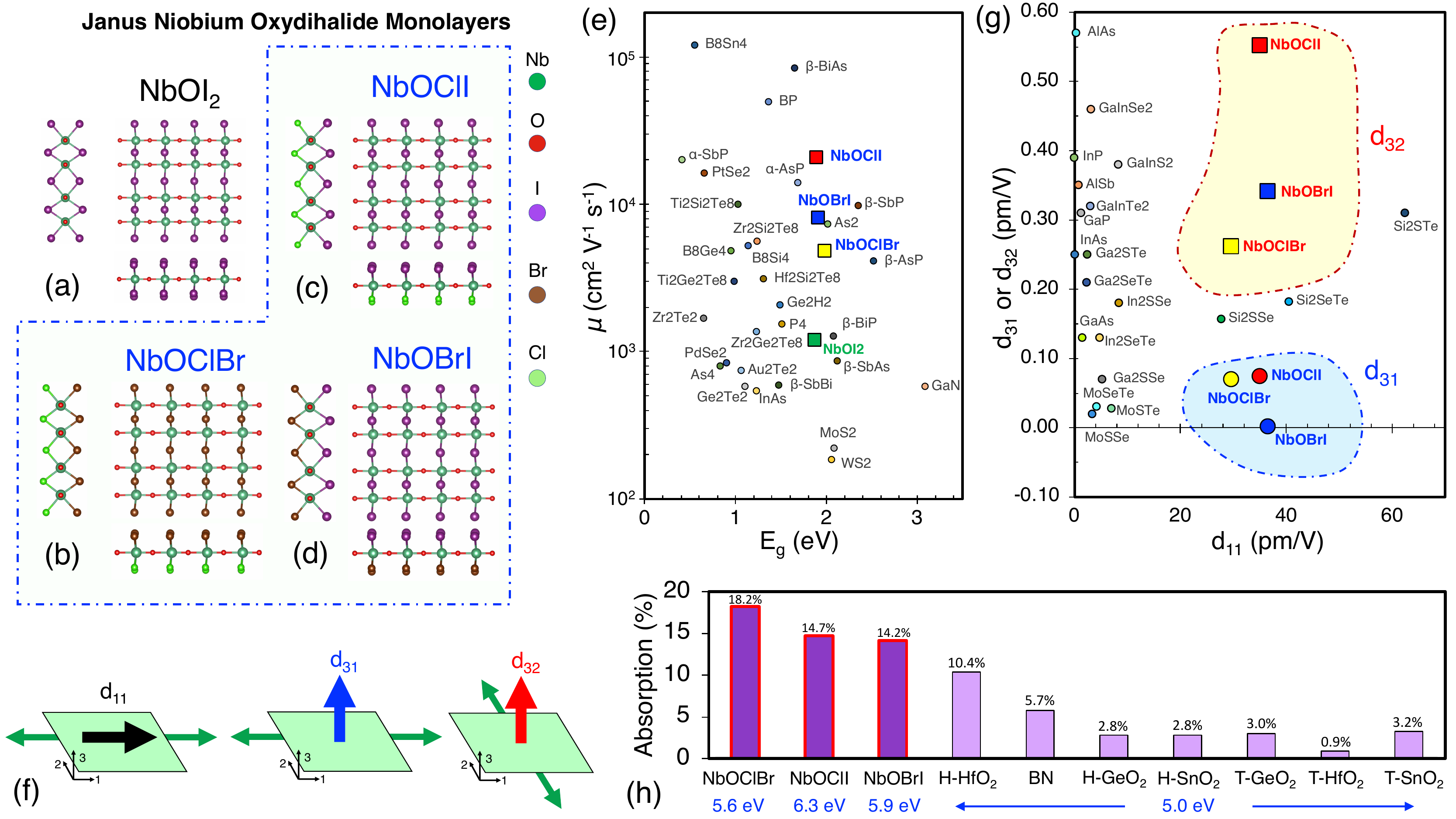}
    \caption{\textbf{Lattice structures and highlights of the electrical, piezoelectronic and optical properties of Janus NbO$XY$.} (a) to (d) shows the lattice structures of non-Janus NbOI$_2$ and the Janus monolayers of NbOClBr, NbOClI, NbOBrI, respectively, in which the inversion symmetry in the out-of-plane direction is broken due to the nonequivalent halogen atoms. (e) Comparison of electrical mobility of various 2D semiconductors obtained from DFT calculations. The NbO$XY$ monolayers mobility is in the range of $10^3 \sim 10^4$ cm$^2$V$^{-1}$s$^{-1}$. (f) Schematic drawings of the $d_{11}$, $d_{31}$ and $d_{32}$ piezoelectric responses. The thin arrows (green) denote the direction of mechanical stress and the thick arrows (black, blue and red) denote the direction of the electric charge polarization ($d_{11}$, $d_{31}$, $d_{32}$). (g) Piezoelectric responses $d_{11}$ and $d_{31}$ (or $d_{32}$) for various 2D materials. (h) Peak optical absorbance of NbO$XY$ in comparison of other 2D materials \cite{2D_abs}. The frequency of the peak absorbance are marked in blue font directly below the monolayer labels. }
    \label{fig:1}
\end{figure*}

Two-dimensional (2D) materials offer an exciting platform for the development of next-generation piezoelectronic technology \cite{cui2018two, li2018recent}. 
The atomically-thin nature of 2D materials and the enormous design space uniquely enabled by van der Waals heterostructure engineering \cite{https://doi.org/10.1002/adma.202109894, pham20222d} offer a new paradigm for designing ultimately-compact and high-performance piezoelectric devices in the \emph{2D Flatland}. 
Myriads of 2D in-plane and out-of-plane piezoelectric materials have been reported recently \cite{cheon2017data}, including TMDCs \cite{duerloo2012intrinsic,wu2014piezoelectricity,zhu2015observation,kim2016directional,Ju2017Reliable,2015Piezoelectric}, hBN \cite{michel2009theory}, graphene nitride \cite{zelisko2014anomalous}), $\alpha$-In$_2$Se$_3$, doped graphene \cite{da2015strong}, and multilayer MoS$_2$ on PbTiO$_3$ \cite{jin2018virtual}). 
The existence 2D materials with \emph{simultaneous} in-plane and out-of-plane piezoelectricity, such as Janus transition metal dichalcogeneides (TMDCs) \cite{lu2017janus,doi:10.1021/acsnano.7b03313}, Janus silicon dichaolcogenide \cite{guo2022two} as well as III-V \cite{chang2014piezoelectric,shi2019electronic} and Janus III-VI monolayers \cite{hinchet2018piezoelectric,guo2017enhanced}, further enriches the application potential of 2D piezoelectric materials. 
The \emph{directionally decoupled} nature of the mechanical stimuli and the corresponding electrical responses is particularly useful due to their compatibility with the stacking design of conventional CMOS and van der Waals heterostructure engineering \cite{liang2020van}. 

Recent high throughput searches \cite{zhou20192dmatpedia,wu2022data} has established niobium oxydihalide monolayers (NbO$X_2$, $X$ = Cl, Br, or I) as an exceptional piezoelectric 2D material family.
Monolayer NbOI$_2$ -- an experimentally \cite{NbOI2_AM, abdelwahab2022giant} fabricated air-stable 2D semiconductors with anisotropic electrical, mechanical and optical properties \cite{jia2019niobium, ye2021ferroelectric, mortazavi2022highly} -- has an exceedingly large piezoelectric response of $d_{11} \approx 45 pm/V$.
Out-of-plane piezoelectricity is, however, strictly forbidden in the NbOX$_2$ monolayer family due to the lattice centrosymmetry, which severely limits the potential of NbOX$_2$ in piezoelectric device applications. 
The absence of out-of-plane piezoelectricity in NbOI$_2$ immediately raises the following questions: Can inversion symmetry breaking from \emph{Janus phase engineering} \cite{lu2017janus,doi:10.1021/acsnano.7b03313} be used as an efficient way to generate sizable out-of-plane and in-plane piezoelectric response? 
Apart from the commonly observed $d_{31}$ piezoelectric response, can other types of out-of-plane piezoelectricity, such as $d_{32}$ response, exist in Janus niobium oxydihalide monolayers? Will the electrical and optical properties of Janus niobium oxydihalide monolayers be beneficial for solid-state device applications, in addition to piezoelectricity?

In this work, we show that Janus-engineered niobium oxyhalide monolayer, NbO$XY$ where $X$ and $Y$ are the halogen atoms Br, Cl, I, and $X\neq Y$, is a multifunctional semiconductor with ultrahigh carrier mobility, mechanical flexibility, strong optical dichroism, broadband visible and strong UV light absorption, compatibility with overall photocatalytic water splitting and simultaneous presence of in-plane and out of-plane piezoelectricity (see Fig. 1 for the crystal structures and a highlight of the electrical, piezoelectric and optical properties of NbO$XY$). 
Using first-principle density functional theory (DFT) calculations, we show that NbO$XY$ is dynamically, thermally and mechanically stable with excellent mechanical flexibility. 
The band gap of 2 eV, which lies in the visible light regime, and the large anisotropic \cite{PhysRevB.102.245419, PhysRevMaterials.4.041001} carrier mobility, ranging from $10^{-1} \sim 10^4$ cm$^2$V$^{-1}$s$^{-1}$ with NbOClI reaching well over $2\times 10^4$ cm$^2$V$^{-1}$s$^{-1}$ which is higher than most 2D semiconductors of similar band gap \cite{sun2021ultrahigh,zhao2017high,banjade2021monolayer,zhang2022discovering,li2022organometallic,xie2016promising,cai2016promising,xiao2018new,xu2018extremely,wang2018electrochemically} [Fig. 1(e)], suggest their strong potential in electronics and optoelectronics applications \cite{qu2022enhanced}. 
The broken inversion symmetry of NbO$XY$ generates both in-plane and out-of plane response -- an uncommon but much sought-after behavior in 2D materials \cite{zhang2021piezotronics,guo2022two}.
Intriguingly, the orthorhombic crystal of MbO$XY$ generates an additional $d_{32}$ response not commonly found in other 2D materials [see Fig. 1(f) for schematic illustrations of $d_{11}$, $d_{31}$ and $d_{32}$ piezoelectric responses]. 
The simultaneously sizable $d_{11}$ and $d_{32}$ responses [see Fig. 1(g) for a comparison with other 2D piezoelectric materials] suggests that NbO$XY$ is an uncommon 2D semiconductor capable of operating in both in-plane and out-of-plane device settings, thus greatly expanding their practicality in piezoelectric and piezoelectronic device applications.
We propose a device concept of \emph{directional stress sensor} to illustrate how the $d_{31}$ and $d_{32}$ of NbO$XY$ responses can be harnessed for complementary piezoelectric functionality not found in their non-Janus counterpart.
$G_0W_0$-BSE calculations further reveals the strong optical anisotropic, broadband visible light absorption, and sharp absorption peaks in the deep UV regime which are much stronger than other 2D optical materials \cite{2D_abs} [Fig. 1(h)], thus unravelling the capability of NbO$XY$ in solar energy harvesting and UV photonics applications \cite{2D_UV_rev}. 
The high-mobility, broadband visible light absorption, presence of built-in electric fields and band edge energies compatible with photocatalytic water splitting \cite{GAN2018352} further suggest the enormous potential of NbO$XY$ for high-efficiency solar-to-hydrogen conversion. 
Our findings unveil NbO$XY$ as a compelling multifunctional semiconductor family with promising potentialin high-performance flexible electronics, optoelectronics, photonics, sensing and sustainable energy harvesting applications.

\section{Computational Methods}

\subsection{Structural Relaxation, Electronic Structures and Transport Properties}
First-principles density functional theory (DFT) \cite{hohenberg1964inhomogeneous} calculations are performed using the projector-augmented wave method as implemented in the Vienna ab initio simulation package (VASP) \cite{kohn1965self, kresse199614251,kresse1994norm}. 
We consider the generalized gradient approximation of Perdew, Burke and Ernzerhof (GGA-PBE) as the exchange-correlation functional \cite{tao2018origin}. 
For band structure calculations, 
we employ the range-separated HSE06 hybrid functionals \cite{heyd2003scuseria,krukau2006influence}.
An energy cut-off of 500 eV, total energy convergence criterion of $10^{-8}$ eV and a force convergence criteria of less than $10^{-3}$ eV/{\AA} on each atom are adopted. 
A vacuum region of $>30$ {\AA} along the $z$ direction is added to avoid interactions between two neighboring images
$\Gamma$-centered $k$-point meshes of $25 \times 13 \times 1$ in the first Brillouin zone is employed to yield well-converged results, i.e. $a$ and $b$ lattice constant, for the unit cells of the Janus monolayers.
VASPKIT is used for postprocessing of the DFT calculation data \cite{wang2021vaspkit}.
Bader charge analysis is performed to obtain the net charge transfer between the atoms \cite{henkelman2006fast}. 

The carrier mobility is calculated based on the deformation potential theory of Bardeen and Shockley \cite{BS},
\begin{equation}
    \mu_{i}^{(\nu)} = \frac{e\hbar^3 C_{\text{2D}, i} }{k_B T m_i^{(\nu)} m_d^{(\nu)} E^{(\nu)2}_{l,i} }
\end{equation}
where the superscript $\nu = e,h$ denotes electron and hole, the subscript $i = x,y$ denotes the two orthogonal directions, $m_i^{(\nu)} = \hbar^{2}\left(\partial^2 \varepsilon_\mathbf{k} / \partial k_i^2\right)$ is the effective mass of $\nu$ carrier, $\varepsilon_\mathbf{k}$ is the energy dispersion around the band edge assuming a parabolic dispersion relation, $\mathbf{k} = (k_x, k_y)$ is the 2D wave vector, $m_d^{(\nu)} = \sqrt{m_x^{(\nu)}m_y^{(\nu)}}$ is the density of states effective mass of $\nu$ carrier, $T$ is the temperature, $C_{\text{2D},i}$ is the 2D elastic modulus along $i$ direction and $E_{l,i}^{(\nu)}$ is the deformation potential constant \cite{BS_DFT}.

\begin{table*}[htbp]
\centering
\label{tab:1}  
\caption{\textbf{Summary of DFT calculation data of NbO$XY$.} Lattice constants $a$ and $b$, monolayer thickness $t$, band gap $E_g$, ionization (VBM) energy of the face with $X$ atoms $E_{ip}^{(X)}$ and with $Y$ atoms $E_{ip}^{(Y)}$, electron affinity (CBM) energy of the surface with $X$ halogen atoms $E_{ea}^{(X)}$ and with $Y$ halogen atoms $E_{ea}^{(Y)}$, built-in dipole potential $\Delta V$, charge transfer from Nb atoms to $X$ halogen atoms $\Delta_{\text{Nb}\to X}$ and to $Y$ halogen atoms to $X$ halogen atoms $\Delta_{\text{Nb}\to Y}$, and the charge differences between the $X$ and $Y$ halogen atoms are listed here.}
\resizebox{\textwidth}{!}{
\begin{tabular}{lcccccccccccc}
\hline\hline\noalign{\smallskip}	
Monolayer & $a$ ({\AA}) & $b$ ({\AA})  & $t$ ({\AA}) 	&	$E_{g}$ (eV) &	$E_{ip}^{(X)}$ (eV)	&	$E_{ip}^{(Y)}$ (eV) &	$E_{ea}^{(X)}$ (eV)	&	$E_{ip}^{(Y)}$ (eV)	&	$\Delta V$ (eV) &	$\Delta_{\text{Nb}\to X}$ (e) &	$\Delta_{\text{Nb}\to Y}$ (e)	&	$\Delta_{X/Y}$ (e) \\
\noalign{\smallskip}\hline\noalign{\smallskip}
NbOI$_2$	&	3.97	&	7.59	&	4.60	&	1.87	&	-6.00	&	-6.00	&	-4.14	&	-4.14	&	0.00	&	0.72	&	0.72	&	0.00\\
NbOClBr	&	3.96	&	6.94	&	4.07	&	1.98	&	-6.86	&	-6.42	&	-4.88	&	-4.44	&	0.44	&	1.09	&	0.91	&	0.17\\
NbOClI	&	3.96	&	7.23	&	4.25	&	1.89	&	-6.87	&	-5.83	&	-4.98	&	-3.94	&	1.04	&	1.10	&	0.70	&	0.41\\
NbOBrI	&	3.96	&	7.36	&	4.41	&	1.91	&	-6.53	&	-5.91	&	-4.62	&	-4.00	&	0.62	&	0.94	&	0.70	&	0.24\\
\noalign{\smallskip}\hline
\end{tabular}
}
\end{table*}

\subsection{Thermal, Dynamical, Mechanical and Piezoelectric Properties Calculations}

To asses the dynamical stability of NbO$XY$ monolayers, we calculate the phonon frequencies using density functional perturbation theory (DFPT) calculations \cite{baroni2001phonons} through the direct supercell method with the $4 \times 2 \times 1$ supercell.
The lattice vector is greater than 15 {\AA} as implemented in the Phonopy code \cite{togo2008first}.
The phonon dispersion at different $q$ points are obtained by computing the force constants on a sufficiently large supercell and Fourier interpolating the dynamical matrices in the primitive cell.
Ab initio molecular dynamics (AIMD) simulations of canonical ensemble (i.e. $NVT$ ensemble) are performed to confirm the thermal stability of NbO$XY$ monolayer at 300 K, for 5 ps with a time step of 1 fs, in a $4 \times 2 \times 1$ supercell configuration. 
The Nose algorithm is used to control the temperature \cite{shuichi1991constant}. 
The piezoelectric effect is determined by third-rank tensors $e_{ijk}$ and $d_{ijk}$. In the contracted Voigt notation, $e_{ijk}$ and $d_{ijk}$ are reduced to $e_{il}$ and $d_{il}$, respectively, where $i,j,k = 1, 2, 3$ denotes the three orthogonal spatial directions, and $l=1,2,\cdots, 6$. The elastic stiffness coefficients($C_{ij}^{\text{3D}}$) are calculated using the finite difference method \cite{wu2005systematic} and the piezoelectric stress tensors ($e_{ij}$) are obtained from DFPT in VASP \cite{baroni2001phonons}, via the following definitions \cite{doi:10.1021/acsnano.7b03313, blonsky2015ab}, 

\begin{subequations}
    \begin{equation}
        \hat{e}_{ijk} = \frac{\partial{P_i}}{\partial\varepsilon_{jk}}=\frac{\partial\sigma_{jk}}{\partial{E_i}},
    \end{equation}
    \begin{equation}
        \hat{d}_{ijk} = \frac{\partial{P_i}}{\partial\sigma_{jk}}=\frac{\partial\varepsilon_{jk}}{\partial{E_i}},
    \end{equation}
\end{subequations}
where $\sigma_{jk}$, $\varepsilon_{jk}$, $P_i$ and $E_i$ are the stress, strain, intrinsic polarization tensors, and macroscopic electric fields, respectively. The piezoelectric strain coefficient, $d_{il}$, can then be determined from $e_{il} = \sum d_{ik} C_{kl}$. 
For monolayers, the 2D elastic coefficients $C_{ij}$ and the piezoelectric stress coefficients $e_{ij}$ are normalized by the slab thickness of the simulation cell along the $z$ direction ($L_z$) via $C_{ij} = L_z C_{ij}^{\text{3D}}$ and $e_{ij} = L_z e_{ij}^{\text{3D}}$, where $C_{ij}^{\text{3D}}$ and $e_{ij}^{\text{3D}}$ are the 3D elastic stiffness and piezoelectric stress coefficients, respectively. Unlike NbOI$_2$ ($mm2$ point group symmetry), NbO$XY$ structures only have $m$ point group symmetry. Consider only the in-plane strain and stress for 2D systems \cite{piezo_theory, de2015database,doi:10.1021/acsnano.7b03313, blonsky2015ab}, we obtain,
\begin{subequations}
    \begin{equation}
        \mathbf{e}^{(\text{2D})} =
        \begin{pmatrix} 
        e_{11} & e_{12} & 0 \\ 
        0 & 0 & e_{23} \\ 
        e_{31} & e_{32} & 0 
        \end{pmatrix},
    \end{equation}
    \begin{equation}
        \mathbf{C}^{(\text{2D})} = 
        \begin{pmatrix} 
        C_{11} & C_{12} & 0 \\ 
        C_{12} & C_{22} & 0 \\ 
        0 & 0 & C_{33}
        \end{pmatrix},
    \end{equation}
    \begin{equation}
        \mathbf{d}^{(\text{2D})}=
        \begin{pmatrix} 
        d_{11} & d_{12} & 0 \\ 
        0 & 0 & d_{23} \\ 
        d_{31} & d_{32} & 0 
        \end{pmatrix}.
    \end{equation}
\end{subequations}
where $d_{11} \neq d_{12}$ and $d_{31} \neq d_{32}$ represent two distinctive in-plane and out-of-plane piezoelectric responses, respectively, and $d_{23}$ is a shear piezoelectric response.  
The elastic properties (i.e. Young's modulus and Poisson's ratio) are averaged by the Voigt-Reuss-Hill schemes \cite{hill1952elastic} and evaluated using the ElasticPOST code (https://github.com/hitliaomq/ElasticPOST) \cite{liao2018alloying,liao2020modeling,jasiukiewicz2008fourth}.

\subsection{Optical Properties Calculations}

For the optical properties, random phase approximation (RPA) \cite{bohm1951collective, ehrenreich1959self} with $G_0W_0$ are employed with $6 \times 6 \times 1$ $k$-point mesh, 300 eV cutoff energy, and the number of bands was tripled. 
Beyond the RPA approach which considers only the dipole transition, single-shot $G_0W_0$ approximation \cite{shishkin2006implementation} is used with the Bethe-Salpeter equation (BSE) \cite{salpeter1951relativistic,onida2002electronic} within the Tamm-Dancoff approximation which captures the electron-hole interactions. 
The complex dielectric function $\varepsilon(\omega)=\varepsilon_1(\omega)+i\varepsilon_2(\omega)$ are computed by considering the electromagnetic wave polarization along both $x$ and $y$ directions. 
The optical absorption coefficient is obtained as \cite{2D_abs},
\begin{equation}
    \alpha(\omega)=\frac{\text{Re}\left[\tilde{\sigma}(\omega)\right]}{\left|1 + \frac{\tilde{\sigma}(\omega)}{2}\right|^2}
\end{equation}
where $\tilde{\sigma}(\omega) \equiv \sigma_{\text{2D}}(\omega)/\varepsilon_0c$, $\sigma_{\text{2D}}(\omega) = i\varepsilon_0\omega L_z\left[ 1  - \varepsilon(\omega )\right]$ is the frequency-dependent complex optical conductivity of a 2D system, $\varepsilon_0$ is the permittivity of free space, $c$ is the speed of light, and $L_z$ is the slab thickness of the simulation supercell. 

\section{Results and Discussions}

\begin{figure*}[t]
    \centering
    \includegraphics[scale=0.7658]{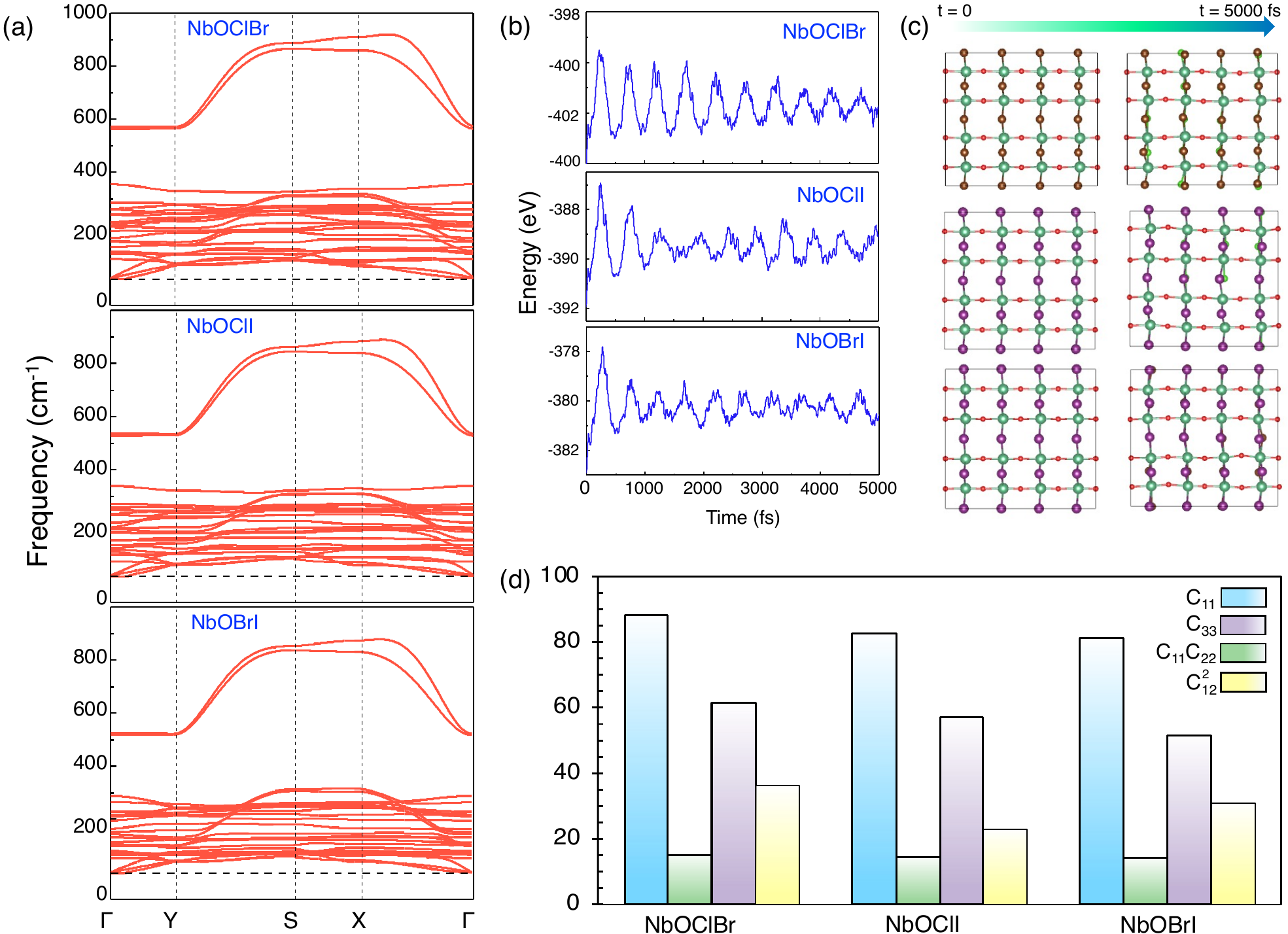}
    \caption{\textbf{Dynamical, thermal and mechanical stability of NbO$XY$ monolayers}. (a) Phonon spectra of the monolayers reveal the absence of significant soft modes around the $\Gamma$ point; (b) Ad initio molecular dynamics (AIMD) simulations of NbO$XY$ monolayers for a duaration of 5000 fs at 300 K. The energy of the 2D systems fluctuate less than 2 meV; (c) Lattice structure of the NbO$XY$ after 5000 ps showing minimal distortions; (d) the elastic coefficients of $C_{11}$, $C_{33}$, $C_{11}C_{12}$ and $C_{12}^2$ of NbO$XY$ fulfil the Born-Huang criteria \cite{Born} of $C_{11}>0$ $C_{33}>0$ and $C_{11}C_{22}>C_{12}^2$. The $C_{11}$ and $C_{33}$ are in the unit of N/m while $C_{11}C_{22}$ and $C_{12}^2$ are in the units of $10^2$ N$^2$/m$^2$ and N$^2$/m$^2$, respectively.}
    \label{fig2}
\end{figure*}

\subsection{Structural Properties and Monolayer Stability}

The lattice structure of NbO$XY$ takes an orthorhombic form with a rectangular network lattice [Fig. 1(a) to (d)]. 
The Janus nature of NbO$XY$ arises from the two outer sublayers of nonequivalent halogen atoms ($X\neq Y$) sandwiching the central Nb atoms. 
The rectangular lattice are highly anisotropic with $x$ and $y$ directions composed of O-Nb-O and X-Nb-Y networks, respectively.
The lattice constants of the fully relaxed NbO$XY$ are 6.93 {\AA}, 7.22 {\AA}, and 7.36 {\AA}, respectively, along the $a$-axis ($x$ direction), and about 3.96 {\AA} along the $b$-axis ($y$ direction) for all Janus monolayers (see Table I for a summary of DFT calculation data). 
The monolayer thicknesses ($t$) are 4.067 {\AA}, 4.250 {\AA} and 4.409 {\AA} for NbOClBr, NbOClI, and NbOBrI, respectively, which are comparable to that of the NbOI$_2$ (4.405 {\AA}). 

We examine the (i) dynamical stability; (ii) thermal stability; and (iii) mechanical stability of Janus NbO$XY$ via phonon spectra calculation, AIMD simulations and elastic coefficients analysis based on the Born-Huang criteria \cite{Born}, respectively (Fig. 2).
As shown in Fig. 2(a), the absence of significant soft modes in the phonon spectra confirms the dynamical stability of NbO$XY$ in the freestanding form.
The AIMD simulations of NbO$XY$ reveals minute energy fluctuation less than 2 meV after 5 ps at 300 K, with only minimal lattice distortion [Figs. 2(b) and 2(c)], thus confirming the thermal stability of NbO$XY$ at room temperature \cite{PhysRevB.103.085422}. 
To achieve mechanical stability, the Born-Huang criteria requires the elastic coefficients to fulfill the inequalities \cite{Born}, $C_{11} > 0$ $C_{33}>0$, and $C_{11}C_{22} > C_{12}^2$. 
All three NbO$XY$ monolayers fulfil the Born-Huang criteria [Fig. 2(d)]. 
The predicted NbO$XY$ monolayers are thus dynamically, thermally and mechanically stable. 
We further note that Janus and non-Janus NbO$XY$ composed of fluorine atoms exhibit significant soft modes (see \textcolor{blue}{Supplementary Materials}), and are thus excluded in this study.

\begin{figure*}[t]
    \centering
    \includegraphics[scale=0.6258]{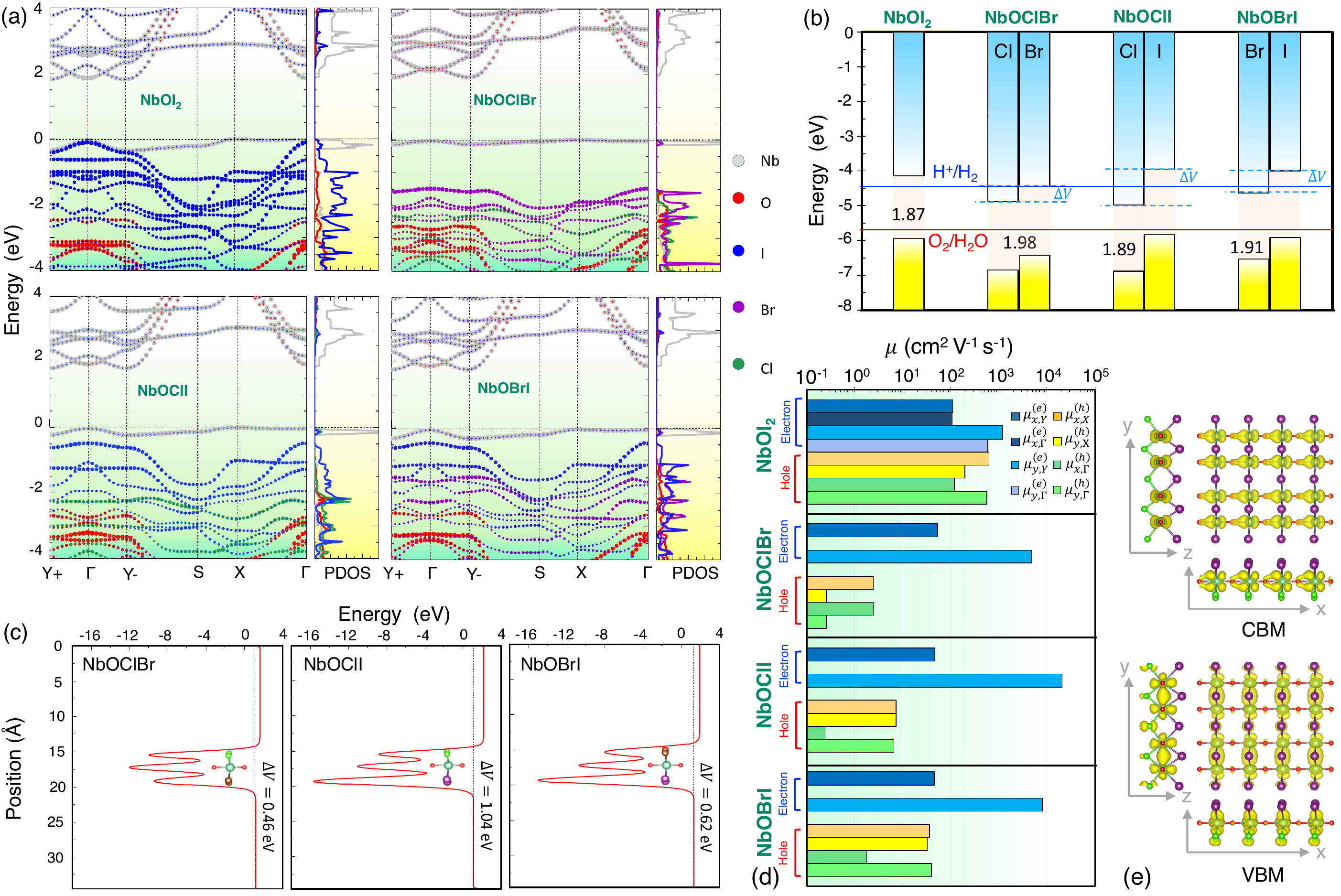}
    \caption{\textbf{Electronic structures of NbO$XY$.} (a) Band structures. (b) The conduction and valance band alignments of NbOI$_2$ and NbO$XY$. The $X$ and $Y$ surface exhibits an offset of $\Delta V$ in Janus NbO$XY$. All values are in the unit of eV. (c) Plane-averaged electrostatic potential plot. (d) Anisotropic electron and hole mobilities of NbO$XY$. (e) The wave function distribution of the CBM and VBM states of NbOClI monolayer.}
    \label{fig3}
\end{figure*}

\subsection{Electronic structures and electrical properties}

\subsubsection{Electronic Band structures} 

The band structures and projected density of states (PDOS) of NbOI$_2$ and NbO$XY$ calcualted via HSE06 method are shown in Fig. 3(a). The band gap is 1.98 , 1.89 and 1.90 eV for NbOClBr, NbOClI and NbOBrI, respectively, which are comparable to that of NbOI$_2$ (1.87 eV) \cite{jia2019niobium}.
Around the CBM, the bands disperses more sharply along the $x$ direction (i.e. $\Gamma \to X$) the dispersion around the $y$ direction (i.e. $\Gamma\to Y_+$ and $\Gamma \to Y_-$) are relatively flatter, which leads to sharply contrasting transport behaviors along these two orthogonal directions.
For non-Janus NbO$X_2$, the electronic states around the conduction band minima (CBM) valance band maxima (VBM) are predominantly from the Nb atoms whereas the VBM also has a sizable contribution from the halogen atoms [see PDOS in Fig. 3(a)] \cite{jia2019niobium}.
Janus NbO$XY$ has a similar electronics structures with CBM and VBM states being dominated by the Nb atoms. 
However, the VBM has a significantly lower contributions from the halogen atoms [see PDOS in Figs. 3(a)].

\subsubsection{Band alignment, interface potential difference and overall photocatalytic water splitting}

The CBM and VBM energies of NbOI$_2$ and NbO$XY$ are shown in Fig. 3(b). 
Interestingly, the CBM (VBM) of NbOI$_2$ is higher (lower) than the water reduction (oxidation) potential of $E_{\text{H}^{+}/\text{H}_2} = -4.44$ eV ($E_{\text{O}^{-}/\text{H}_2\text{O}} = -5.67$ eV) \cite{wang20222d,ji2018janus} for hydrogen (oxygen) molecule production, thus suggesting the potential of NbOI$_2$ in photocatalytic overall water splitting application \cite{obeid2020first}. 
Due to the inversion symmetry breaking of Janus NbO$XY$, unequal charge transfer to the $X$ and $Y$ halogen atoms [see Table I for a summary for the Barder charge analysis] leads to a built-in intrinsic electric field along the out-of-direction of the lattice \cite{zhang_MTP, YUAN2021100549}. 
Correspondingly, a sizable interface potential difference ($\Delta V$) arises across the two surfaces of NbO$XY$, causing the CBM and VBM to offset by $\Delta V$ on the $X$ and $Y$ planar surface as illustrated in the band alignment diagram in Fig. 3(b) as well as the plane-averaged electrostatic potential profile in Fig. 3(c). 
Here $\Delta V$ signifies a work function difference on the two opposite planar surface of NbO$XY$ and is directly proportional to the charge transfer difference between Nb$\to$X and Nb$\to$Y (see \textcolor{blue}{Supplementary Materials}).
While NbOClBr is incompatible for hydrogen evolution reaction (HER) due to the close proximity of the CBM to $E_{\text{H}^{+}/\text{H}_2}$, NbOClI and NbOBrI are compatible with both HER and oxygen evolution reaction (OER) with sizable energy differences, $\Delta E_1 \equiv E_{\text{CBM}} - E_{\text{H}^{+}/\text{H}_2}$ and $\Delta E_2 \equiv E_{\text{O}^{-}/\text{H}_2\text{O}} - E_{\text{VBM}}$, which are critically needed for enhancing the HER and OER activities \cite{wang20222d}. 
The $(\Delta E_1, \Delta E_2)$ of NbOClI and NbOBrI are $(0.50, 1.20)$ eV and $(0.44, 0.86)$ eV, respectively, which significantly outperforms that of the non-Janus NbOI$_2$, i.e. $(\Delta E_1, \Delta E_2) = (0.30, 0.28)$ eV. 
In addition, the presence of a built-in electric field in NbO$XY$ enables the efficient spatial separation of photoexcited electron-hole pairs onto the two opposite surface \cite{ji2018janus}, with electron (hole) preferentially migrates towards the surface with $X$ ($Y$) halogen atoms upon photoexcitation. 
Importantly, the combined factors of (i) sizable $(\Delta E_1, \Delta E_2)$ that is beneficial for enhancing HER and OER activities; (ii) built-in electric field that is beneficial for photoexcited electron-hole pair separation; (iii) strong optical absorption covering visible and ultraviolet regimes that is beneficial for optimal coverage of solar spectrum (discussed below); and (iv) large carrier mobility reaching well over $10^4$ cm$^2$V$^{-1}$s$^{-1}$ (discussed below), suggest the capability of NbO$XY$ monolayers in photocatalytic water splitting applications \cite{CHONG2022101636, LV2022100191}. 

\begin{table*}
\caption{\textbf{Electrical tranpsort properties of NbOI$_2$ and NbO$XY$.} The effective masses, $m_{i}^{(\nu)}$, are in the unit of $m_0$ where $m_0$ is free electron mass. The deformation potential constants, $E_{l,i}^{(\nu)}$, are in the unit of eV. The elastic moduli, $C_{\text{2D},i}$, are in the unit of Jm$^{-2}$. The carrier mobilities, $\mu_{i}^{(\nu)}$, are in the unit of cm$^2$V$^{-1}$s$^{-1}$. The band edge position, i.e. $\Gamma$, $Y_{\pm}$ and $X$, are marked directly after the numeric values as superscript for simplicity. }
\resizebox{\linewidth}{!}{ 
\begin{tabular}{lcccccccccccccccccc}
\hline\hline\noalign{\smallskip}	
Material	&	$m_x^{(e)}$	&	$m_y^{(e)}$ &	$m_x^{(h)}$	&	$m_y^{(h)}$ &	$E_{l,x}^{(e)}$ 	&	$E_{l,y}^{(e)}$ &	$E_{l,x}^{(h)}$ 	&	$E_{l,y}^{(h)}$	&	$C_{\text{2D},x}$	&	$C_{\text{2D},y}$	&	$\mu_x^{(e)}$	&	$\mu_y^{(e)}$ &	$\mu_x^{(h)}$	&	$\mu_y^{(h)}$ \\
\noalign{\smallskip}\hline\noalign{\smallskip}
NbOI$_2$	&	$0.916^{Y_\pm}$	&	$0.797^{Y_\pm}$	&	$8.082^X$	&	$1.219^X$ &	$4.40^{Y_\pm}$	&	$1.81^{Y_\pm}$	&		$0.32^X$	&	$1.33^X$	&	75.61	&	61.87	&	$106.609^{Y_\pm}$	&	$593.647^{Y_\pm}$	&	$621.010^X$	&	$195.035^X$\\
&	$0.334^\Gamma$	&	$0.622^\Gamma$	&	$33.688^\Gamma$	&	$0.782^\Gamma$	&	$9.85^\Gamma$	&	$1.97^\Gamma$	&	$0.28^\Gamma$	&	$0.76^\Gamma$	&&&	$109.217^\Gamma$	&	$1199.744^\Gamma$	&	$119.213^\Gamma$	&	$563.216^\Gamma$\\
NbOClBr	&	$0.500^{Y_\pm}$	&	$1.370^{Y_\pm}$	&	$13.010^X$	&	$30.540^X$ 	&	$9.50^{Y_\pm}$	&	$0.50^{Y_\pm}$	&	$1.78^X$	&	$2.99^X$	&	92.31	&	64.26	&	$52.733^{Y_\pm}$	&	$4836.536^{Y_\pm}$	&	$2.397^X$	&	$0.252^X$\\
&&&	$168.440^\Gamma$	&	$29.218^\Gamma$	&&&	$1.87^\Gamma$	&	$3.67^\Gamma$	&&&&&	$0.048^\Gamma$	&	$0.050^\Gamma$\\
NbOClI	&	$0.578^{Y_\pm}$	&	$1.090^{Y_\pm}$	&	$11.395^X$	&	$6.095^X$ 	&	$9.30^{Y_\pm}$	&	$0.28^{Y_\pm}$	&	$1.64^X$	&	$1.96^X$	&	85.83	&	66.41	&	$46.150^{Y_\pm}$	&	$20888.791^{Y_\pm}$	&	$7.170^X$	&	$7.261^X$\\
&&&	$128.701^\Gamma$	&	$0.28^\Gamma$	&&&	$1.79^\Gamma$	&	$2.12^\Gamma$	&&&&&	$0.240^\Gamma$	&	$6.457^\Gamma$\\
NbOBrI	&	$0.614^{Y_\pm}$	&	$1.014^{Y_\pm}$	&	$9.989^X$	&	$2.410^X$ 	&	$9.01^{Y_\pm}$	&	$0.46^{Y_\pm}$	&	$1.00^X$	&	$1.88^X$	&	83.17	&	63.40	&	$45.118^{Y_\pm}$	&	$8094.971^{Y_\pm}$	&	$36.206^X$	&	$32.366^X$\\
&&&	$71.119^\Gamma$	&	$1.380^\Gamma$	&&&	$1.19^\Gamma$	&	$1.59^\Gamma$	&&&&&	$1.770^\Gamma$	&	$39.153^\Gamma$\\
\noalign{\smallskip}
\hline
\end{tabular}
}
\end{table*}

\subsubsection{Transport properties}

The electrical mobilities of NbOI$_2$ and NbO$XY$ are calculated at room temperature ($T = 300$ K). [see Fig. 3(d) and Table II].
From the deformation potential theory of Bardeen and Shockley \cite{BS} in Eq. (1), the electrical mobility follows the proportionality, $\mu_{i}^{(\nu)} \propto C_{2D,i} / m_{i}^{(\nu)-3/2} E_{l,i}^{(\nu)-2}$. As $\mu$ has a second order dependence on $E_l$, which is sensitively influenced by the wave function distribution of the band edge states, the spatial extend of the band edge states thus play an important role in governing carrier transport.
We first consider electron mobility in the conduction band. 
For NbOI$_2$, as the conduction band edge at the $\Gamma$ and the $Y_\pm$ points are separated by only about 26 meV, both \emph{valleys} are expected to contribute to the electron conduction current. The electron mobilities of both valleys are calculated, i.e. $\mu^{(e)}_{i, Y_\pm}$ and $\mu^{(e)}_{i,\Gamma}$, respectively, where $i = x,y$ represents the two orthogonal crystal directions. 
In contrast, the $\Gamma$ valley in NbO$XY$ is energetically well-separated from the actual CBM at $Y_{\pm}$ valley by a substantial energy, we thus consider only $\mu^{(e)}_{i, Y_\pm}$ for NbO$XY$.
The electron mobility is highly anisotropic in both NbOX$_2$ and NbO$XY$ with $\mu_{x, Y_\pm/\Gamma}^{(e)} \ll \mu_{y, Y_\pm/\Gamma}^{(e)}$ [Fig. 3(d)].
The larger electron mobility in the $y$ direction ($\mu_{y, Y_\pm/\Gamma}^{(e)}$) is a direct consequence of the weak orbital overlap along the $y$ crystal direction [see the band-decomposed charge density distributions of the CBM states of NbOClI as a representative example in Fig. 3(e)]. 
Lattice distortion along the $y$ direction does not generate significant CBM energy shifting, thus resulting in small $E^{(\text{CBM})}_{l,y}$. 
Since $\mu_y^{(e)} \propto \left(E^{(\text{CBM})}_{l,y}\right)^{-2}$, the electron mobility is significantly enhanced, yielding high mobility values of $\mu_{y, Y_\pm/\Gamma}^{(e)} = 10^3 \sim 10^4 \times 10^4$ cm$^2$V$^{-1}$s$^{-1}$ in NbOXY. 
Particularly, NbOClI has an exceedingly large mobility $\mu_{y, Y_\pm}^{(e)} = 2 \times 10^4$ cm$^2$V$^{-1}$s$^{-1}$, which is considerably higher than the vast majority of 2D semiconductors with band gap $E_g \gtrsim 2$ eV.

For the hole conduction in valance bands, because of the close energetic proximity of the $\Gamma$ point and the CBM at $X$ point with minute energy separation $\sim 40$ meV, we calculate the hole mobility at both $X$ and $\Gamma$ point, i.e. $\mu^{(h)}_{i, X}$ and $\mu^{(h)}_{i,\Gamma}$, for all monolayers.
Interestingly, the wave function distribution of the VBM states is contrary to that of the CBM states, i.e. strong orbital overlap strongly (weakly) along the $x$ ($y$) crystal direction [see Fig. 3(e)]. 
However, for hole conduction at the $X$ point of NbOClI and NbOBrI, the anisotropy of $E_{l,x}<E_{l,y}$ is \emph{compensated} by a sizable hole effective mass anisotropy of $m_x^{(h)} > m_y^{(h)}$. 
Their hole mobilities are thus comparable or approximately isotropic at the $X$ points, i.e. $\mu^{(h)}_{x,X} \approx \mu^{(h)}_{x,X}$. 
At the $\Gamma$ point of NbOClI and NbOBrI, the anisotropy of $E_{l,x}<E_{l,y}$ is insufficient to negate the exceedingly large hole effective mass of $m_{x}^{(h)} \gg m_{y}^{(h)}$. 
An unexpected mobility anisotropy of $\mu^{(h)}_{x,\Gamma} \ll \mu^{(h)}_{x,\Gamma}$ thus occurs despite the weak orbital overlap along the $x$ crystal direction. 

It should be noted that the electron mobility anisotropy $\mu_{x,Y_\pm/\Gamma}^{(e)} < \mu_{y,Y_\pm/\Gamma}^{(e)}$ and the hole conduction anisotropy at $\Gamma$ point $\mu^{(h)}_{x,\Gamma} < \mu^{(h)}_{x,\Gamma}$ originate from completely different mechanisms: (i) $\mu^{(h)}_{x,\Gamma} < \mu^{(h)}_{x,\Gamma}$ is caused by the extraordinarily strong hole effective mass anisotropy; (ii) whereas $\mu_{x,Y_\pm/\Gamma}^{(e)} < \mu_{y,Y_\pm/\Gamma}^{(e)}$ arises from the nearly absent $y$-directional wave function overlap of the CBM states. 
We further note that, hole transport in NbOClBr is distinctive from the other NbO$XY$ monolayers.
NbOClBr exhibits $\mu^{(h)}_{x,X/\Gamma} > \mu^{(h)}_{y,X/\Gamma}$ at both $X$ and $\Gamma$ points due to the distinctive hole effective mass anisotropy: 
(i) $m_x^{(h)} < m_y^{(h)}$ at the $X$ point and $E_{l,x}^{(h)} < E_{l,y}^{(h)}$ jointly enhances $x$-directional hole mobility, leading to $\mu^{(h)}_{x,X} > \mu^{(h)}_{y,X} $; whereas (ii) the milder (and opposite) effective mass anisotropy of $m_x^{(h)} > m_y^{(h)}$ at the $\Gamma$ point is insufficient to negate the $E_{l,x}^{(h)} < E_{l,y}^{(h)}$ that amplifies $x$-directional transport, thus leading to $\mu^{(h)}_{x,\Gamma} > \mu^{(h)}_{y,\Gamma}$ which is akin to the case of $X$ point.

\begin{figure*}[t]
    \centering
    \includegraphics[scale=0.4558]{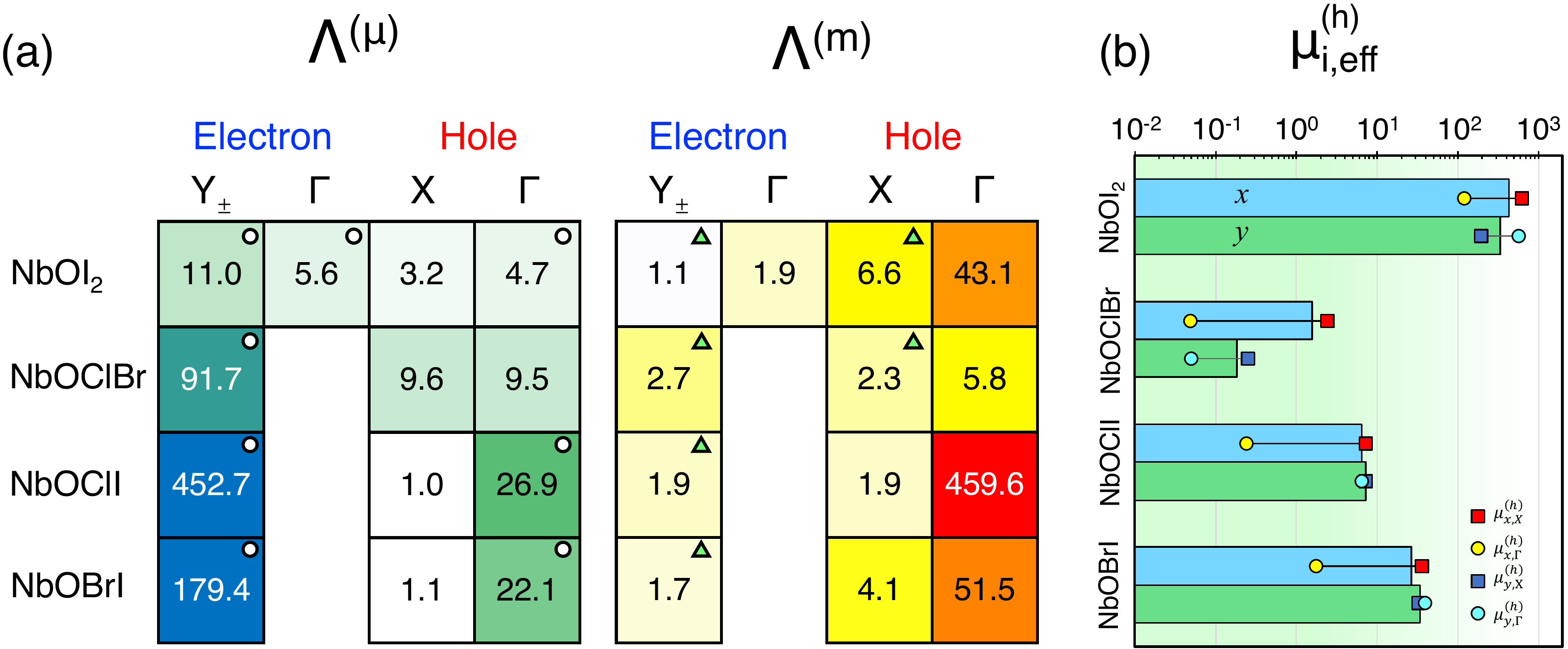}
    \caption{\textbf{Transport anisotropy and effective hole mobility.} (a) shows the electrical mobility ratios and effective mass ratios at different conduction and valance band valleys. The circle and triangle symbols denote $\mu_{x}^{(\nu)} < \mu_{y}^{(\nu)}$ and $m_x^{(\nu)} < m_y^{(\nu)}$, respectively. Otherwise, $\mu_{x}^{(\nu)} > \mu_{y}^{(\nu)}$ and $m_x^{(\nu)} > m_y^{(\nu)}$. (b) Effective hole mobility of multi-valley hole transport in NbOI$_2$ and NbO$XY$. Blue and green bars denote $x$-and $y$-directional hole effective mobility, respectively. }
    \label{fig5}
\end{figure*}

\subsubsection{Transport anisotropy and effective mobility}

We summarize the mobility and effective mass anisotropy in Fig. 4(a) where the anisotropy ratios are defined as,
\begin{subequations}
    \begin{equation}
        \Lambda^{(\mu)} \equiv \frac{\text{max}\left[\mu^{(\nu)}_{x}, \mu^{(\nu)}_{y} \right]}{ \text{min}\left[\mu^{(\nu)}_{x}, \mu^{(\nu)}_{y} \right] } ,
    \end{equation}
    \begin{equation}
        \Lambda^{(m)} \equiv \frac{ \text{max}\left[ m^{(\nu)}_x, m^{(\nu)}_y \right] }{ \text{min}\left[ m^{(\nu)}_x, m^{(\nu)}_y \right] },
    \end{equation}
\end{subequations}
$\text{max}[\cdots]$ and $\text{min}[\cdots]$ are the maximum and minimum functions, respectively. The electron mobility of NbO$XY$ exhibits the enormous mobility anisotropy in the order of $10^2$ whereas the $\Gamma$-point hole effective mass exhibits strong anisotropy in the order of $10 \sim 10^{2}$. 
Such exceptionally anisotropic transport properties of NbO$XY$ suggests new avenues for nanoelectronics device engineering in which the carrier mobility can be tuned over two orders of magnitude by selectively aligning the electrodes along different crystal directions.

\begin{figure*}[t]
    \centering
    \includegraphics[scale=0.5858]{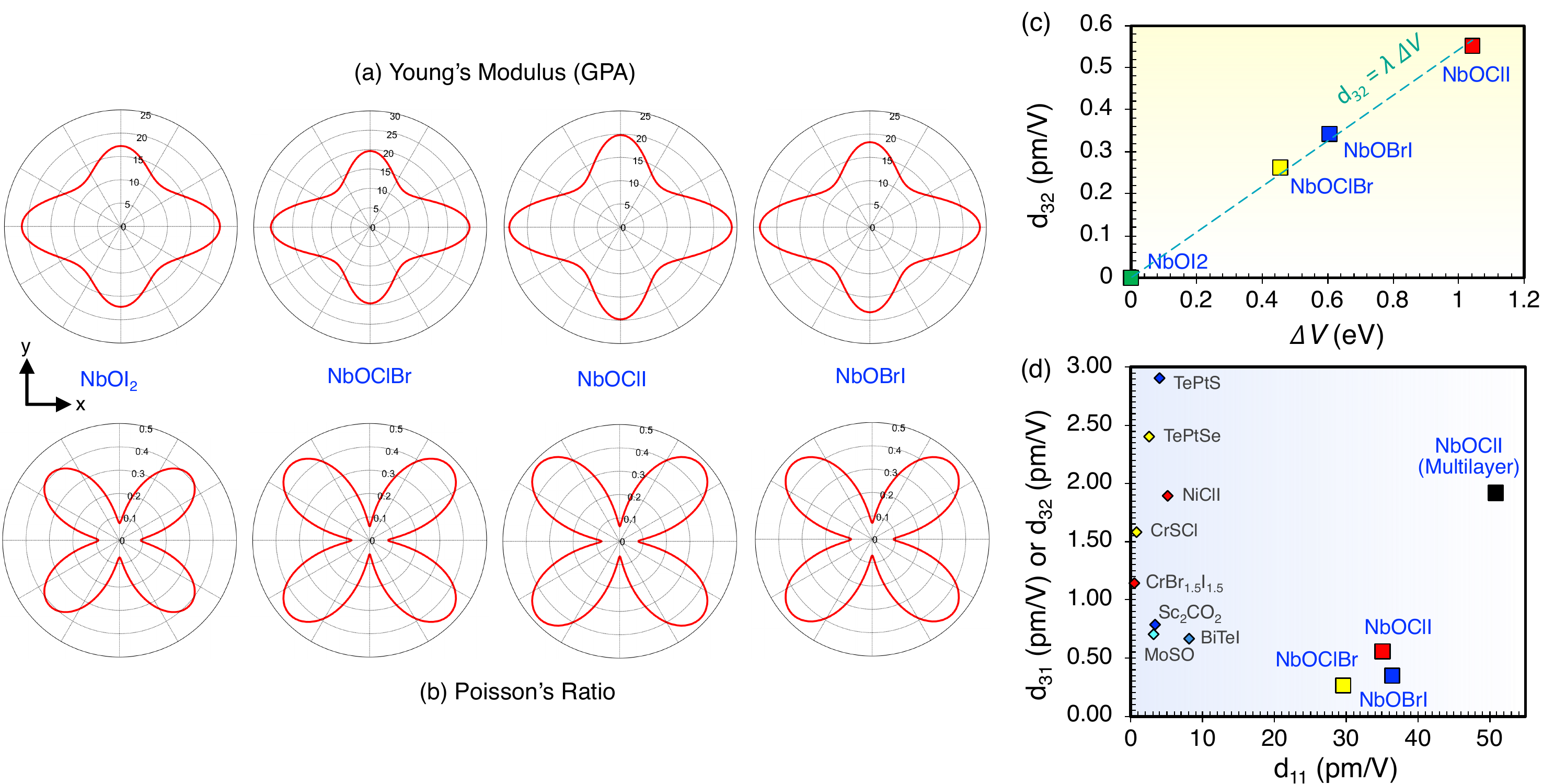}
    \caption{\textbf{Anisotropic elastic properties and out-of-plane piezoelectric response of NbO$XY$ monolayers}. Directional dependence of (a) Young's modulus and (b) Poisson's Ratio. (c) Correlation between $d_{32}$ and work function differences $\Delta V$. The fitting equation $d_{32} = \lambda \Delta V$ has an excellent linear fit with $R^2$ value of 0.999. (d) Comparison of NbO$XY$ ($d_{11}$ and $d_{32}$) with other 2D materials of sizable out-of-plane piezoelectricity (i.e. $d_{32} > 0.5$ pm/V). The in-plane and out-of-plane piezoelectric responses of multilayer NbOClI is significantly larger than the monolayer counterpart.}
    \label{fig4}
\end{figure*}

The presence of multiple valance band valleys in NbO$XY$ can co-contribute to hole conduction current. 
However, the electrical current from different valleys cannot be directly distinguished from typical transport measurement. 
It is thus important to determine the \emph{effective hole mobility} of the total transport current that includes the current conduction in multiple valleys \cite{herring1955transport, 1476105}. 
We consider a drift current transport picture in which the total hole current density is given by
\begin{equation}
    \mathcal{J}_{i, \text{T}}^{(h)} = e n_{\text{T}}^{(h)} \mu_{i, \text{eff}}^{(h)} \mathcal{E},
\end{equation}
where $\mathcal{E}$ is the applied electric field along the $i = x,y$ direction, $\mu_{i,\text{(eff)}}^{(h)}$ is the effective hole mobility inclusive of both $X$ and $\Gamma$ valleys, and $n_{\text{T}}^{(h)} = n_{X}^{(h)} + n_{\Gamma}^{(h)}$ is the \emph{total} 2D carrier density inclusive of both $X$ and $\Gamma$ valleys, i.e. $n_{X}^{(h)}$ and $n_{\Gamma}^{(h)}$, respectively. 
Since the total hole current density is contributed by both valleys, we can equivalently write
\begin{equation}
    \mathcal{J}_{i, \text{T}}^{(h)} = \mathcal{J}_{i,X}^{(h)} + \mathcal{J}_{i,\Gamma}^{(h)},
\end{equation}
where $\mathcal{J}_{i,X}^{(h)} = e n_{X}^{(h)} \mu_{i,X}^{(h)} \mathcal{E}$ and $\mathcal{J}_{i,\Gamma}^{(h)}= e n_{\Gamma}^{(h)} \mu_{i,\Gamma}^{(h)} \mathcal{E}$ are the $X$ and $\Gamma$ valley current component, respectively. 
The 2D carrier densities are
\begin{equation}
    n^{(h)}_{\lambda} = \frac{2}{(2\pi)^2} \int_{-\infty}^{-\varepsilon_\lambda} \text{d}k_x \text{d}k_y \left[ 1 - f\left( \varepsilon_\mathbf{k}, \varepsilon_F,T \right) \right],
\end{equation}
where $\lambda = X, \Gamma$ denotes the band edge position, $f\left( \varepsilon_\mathbf{k}, \varepsilon_F,T \right)$ is the Fermi-Dirac distribution function and $\varepsilon_F$ is the Fermi level. Consider the non-degenerate regime with $\varepsilon_F$ several times larger than $k_BT$ which is well achievable at room temperature, the Fermi-Dirac distribution function reduces to the semiclassical limit of $f\left( \varepsilon_\mathbf{k}, \varepsilon_F,T \right) \approx \exp\left[ - \left( \varepsilon_\mathbf{k} - \varepsilon_F\right) / k_BT \right]$. 
Equations (6) to (8) can be combined to yield
\begin{equation}
    \mu_{i,\text{eff}}^{(h)} = \frac{\mu_{i,X}^{(h)}}{ 1 + \frac{n^{(h)}_{\Gamma}}{ n^{(h)}_{X} } } \left( 1 + \frac{n^{(h)}_{\Gamma}}{ n^{(h)}_{X} } \frac{ \mu^{(h)}_{i,\Gamma} }{ \mu^{(h)}_{i,X} } \right),
\end{equation}
where the hole density ratio is
\begin{equation}
    \frac{n^{(h)}_{\Gamma}}{ n^{(h)}_{X} } = \frac{ m_{d,\Gamma}^{(h)}}{ m_{d,\Gamma}^{(h)} } \exp\left( - \frac{\Delta_{X-\Gamma}}{k_BT} \right),
\end{equation}
Here $m_{d,\lambda}^{(h)} = \sqrt{m_{x,\lambda}^{(h)} m_{y,\lambda}^{(h)} }$ is the DOS effective mass of hole at the $\lambda$ point and $\Delta_{X-\Gamma} \equiv \left| \varepsilon_X - \varepsilon_\Gamma \right|$ is the energy separation between $X$ and $\Gamma$ points. 
The effective hole mobility along the $x$ and $y$ directions are shown in Fig. 4(b). 
As $\mu_{i,\text{eff}}^{(h)}$ in Eq. (9) is essentially the carrier density-weighted average of $\mu^{(h)}_{i,X}$ and $\mu^{(h)}_{i,\Gamma}$, the effective mobility $\mu_{i,\text{eff}}^{(h)}$ is intermediate between the $\mu^{(h)}_{i,X}$ and $\mu^{(h)}_{i,\Gamma}$. 
Since $\mu_{i,\text{eff}}^{(h)}$ takes into account the multi-valley transport of holes, we expect $\mu_{i,\text{eff}}^{(h)}$ to exhibit a better agreement with experimental transport measurement when compared with the individual valley mobilities of $\mu^{(h)}_{i,X}$ and $\mu^{(h)}_{i,\Gamma}$. 
In general, $\mu_{x,\text{eff}}^{(h)} \approx \mu_{y}, \text{eff}^{(h)}$ for NbOI$_2$, NbOCl and NbOBrI. In contrast, NbOClBr exhibits a more dramatic hole effective mobility anisotropy due to the effective mass anisotropy of $m_{x}^{(h)} < m_{y}^{(h)}$ at $X$ point as discussed above. 

\begin{table*}[htbp]
\centering
\label{tab:3}  
\caption{\textbf{Mechanical and piezoelectric properties of NbO$XY$.} The piezoelectric stress coefficients $e_{ij}$ in the unit of $10^{-10}$ C/m, elastic stiffness coefficients $C_{ij}$ in the unit of $N/m$, and the piezoelectric strain coefficients $d_{ij}$ in the unit of pm/V. }
\begingroup
\setlength{\tabcolsep}{8pt} 
\renewcommand{\arraystretch}{1.05} 
\begin{tabular}{lcccccccccccc}
\hline\hline\noalign{\smallskip}	
Monolayer	&	$e_{11}$ &	$e_{12}$ &	$e_{31}$ &	$e_{32}$ &	$e_{23}$ &	$C_{11}$	&	$C_{12}$	&	$d_{11}$ &	$d_{12}$ &	$d_{31}$ &	$d_{32}$ &	$d_{23}$ \\
\noalign{\smallskip}\hline\noalign{\smallskip}
NbOI$_2$\footnote{DFT calculation data from Ref. \cite{wu2022data}}	&	31.60	&	-1.00	&	0.00	&	0.00	&	0.70	&	75.60	&	5.30	&	42.20	&	-5.10	&	0.000	&	0.00	&	5.20\\
NbOI$_2$	&	31.70	&	-1.11	&	0.00	&	0.00	&	0.71	&	73.31	&	4.91	&	43.60	&	-5.40	&	0.000	&	0.00	&	5.21\\
NbOClBr	&	25.91	&	-1.29	&	0.07	&	0.19	&	0.86	&	88.11	&	6.01	&	29.62	&	-3.90	&	0.070	&	0.26	&	5.80\\
NbOClI	&	28.76	&	-1.28	&	0.09	&	0.38	&	0.90	&	82.66	&	4.80	&	35.05	&	-4.37	&	0.074	&	0.55	&	6.29\\
NbOBrI	&	29.34	&	-1.22	&	0.02	&	0.22	&	0.75	&	81.18	&	5.57	&	36.49	&	-5.11	&	0.002	&	0.34	&	5.31\\
\noalign{\smallskip}
\hline
\end{tabular}
\endgroup
\end{table*}

\subsection{Mechanical and Piezoelectric Properties}

\subsubsection{Young's modulus and Poisson's ratio}

We now examine the Young's modulus and Poisson's ratio of NbO$XY$(Fig. 5). 
Table III summarizes the mechancial properties of NbO$XY$.
NbO$XY$ exhibits mechanical anisotropy in terms of Young's modulus [Fig. 5(a)]. 
The Young's modulus peaks along $x$ and $y$ directions, with values lying between $20$ GPa and $25$ GPa along the $x$ direction, and between $18$ GPa and 20 GPa along the $y$ direction. 
The anisotropy ratio of the Young's modulus is only $\lesssim 2$ for NbO$XY$, which is less profound when compared to other 2D materials with giant mechanical anisotropy \cite{zhao2021recent,puebla2021plane}.
It is noteworthy that the Young's modulus of NbO$XY$ is significantly smaller than many other 2D materials, such as graphene, boron nitride and TMDC \cite{jiang2020two}, thus suggesting their excellent mechanical flexibility useful for flexible electronics applications. 
Unlike auxetic 2D materials \cite{peng2019two}, the Poisson's ratio of NbO$XY$ is entirely positive and exhibits directional dependence with maximum values of $\sim 0.45$ along the directions 45$^\circ$ to the $x$ and $y$ axis [Fig. 5(b)].

\subsubsection{Piezoelectric coefficients}

The 2D piezoelectric stress coefficients $e_{ij}$ and the piezoelectric strain coefficients $d_{ij}$ are summarized in Table III. 
In contrast to the centrosymmetric NbOI$_2$ with $d_{31} = 0$, the broken inversion symmetry in NbO$XY$ generates two distinctive out-of-plane piezoelectric responses, as characterized by $d_{31}$ and $d_{32}$. 
Janus NbO$XY$ monolayers retain the large $d_{11}$ of 35.05, 29.62 and 36.49 pm/V for NbOClI, NbOClBr and NbOBrI, respectively.
Such $d_{11}$ values are only slightly lower than that of NbOI$_2$ and outperforms a large number of 2D in-plane piezoelectric materials [Fig. 1(g)]. 
The out-of-plane $d_{31}$ responses are 0.074, 0.070 and 0.002 pm/V for NbOClI, NbOClBr and NbOBrI, respectively. 
The $d_{31}$ of NbOClI and NbOClBr are more than 2 times larger than that of MoSTe monolayer (0.030 pm/V) -- the top performer in the Janus TMDC family \cite{doi:10.1021/acsnano.7b03313} -- and several other 2D out-of-plane piezoelectric materials [Fig. 1(g)]. 
In addition, the $d_{32}$ coefficient is several times larger than $d_{31}$, with 0.26, 0.55 and 0.34 pm/V for NbOClI, NbOClBr and NbOBrI, respectively. 
Here NbOClI is a particularly exceptional 2D semiconductor with simultaneously large in-plane ($d_{11}$) and sizable out-of-plane ($d_{32}$) piezoelectricity. 
Interestingly, $d_{32}$ is directly proportional to the interface potential difference $\Delta V$, exhibiting an excellent linear fit (i.e. $R^2 \approx 0.999$) [Fig. 5(c)] which can be captured by a simple semi-empirical expression,  
\begin{equation}
    d_{32} = \lambda \Delta V,
\end{equation}
where $\lambda = 0.5421$ pm/eV$^2$. 
Here $d_{32}$ and $\Delta V$ are in the units of pm/V and eV, respectively. 
The linear relationship between out-of-plane piezoelectricity in 2D Janus materials and $\Delta V$ has not been reported previously. 
Such linear relationship suggests the importance of having a strong built-in dipole potential in order to achieve strong $d_{32}$ responses.
It should be noted that although other 2D piezoelectric materials have been predicted to exhibit out-of-plane piezoelectric responses stronger than NbO$XY$, most of these materials have significantly smaller $d_{11} < 10$ pm/V. 
Figure 5(d) compares NbO$XY$ with other 2D piezoelectric materials with large out-of-plane response, i.e. $d_{31} > 0.5$ pm/V, in terms of their in-plane and out-of-plane piezoelectricity.
It can be seen that $d_{31}$ or $d_{32}$ generally decreases with larger $d_{11}$.
Nevertheless, NbOClI still retains a sizable out-of-plane piezoelectricity of $d_{32} = 0.55$ pm/V despite having an exceedingly large $d_{11} = 35.05$ pm/V.
We further calculate the piezoelectric coefficients of \emph{multilayer} NbOClI, which yields significantly amplified piezoelectric coefficients of $d_{11} = 50.86$ pm/V, $d_{31} = 1.92$ pm/V and $d_{32} = -1.04$ pm/V, thus suggesting that layer number engineering can be used to effectively enhance both in-plane and out-of-plane piezoelectricity.

\begin{figure*}[t]
    \centering
    \includegraphics[scale=0.5658]{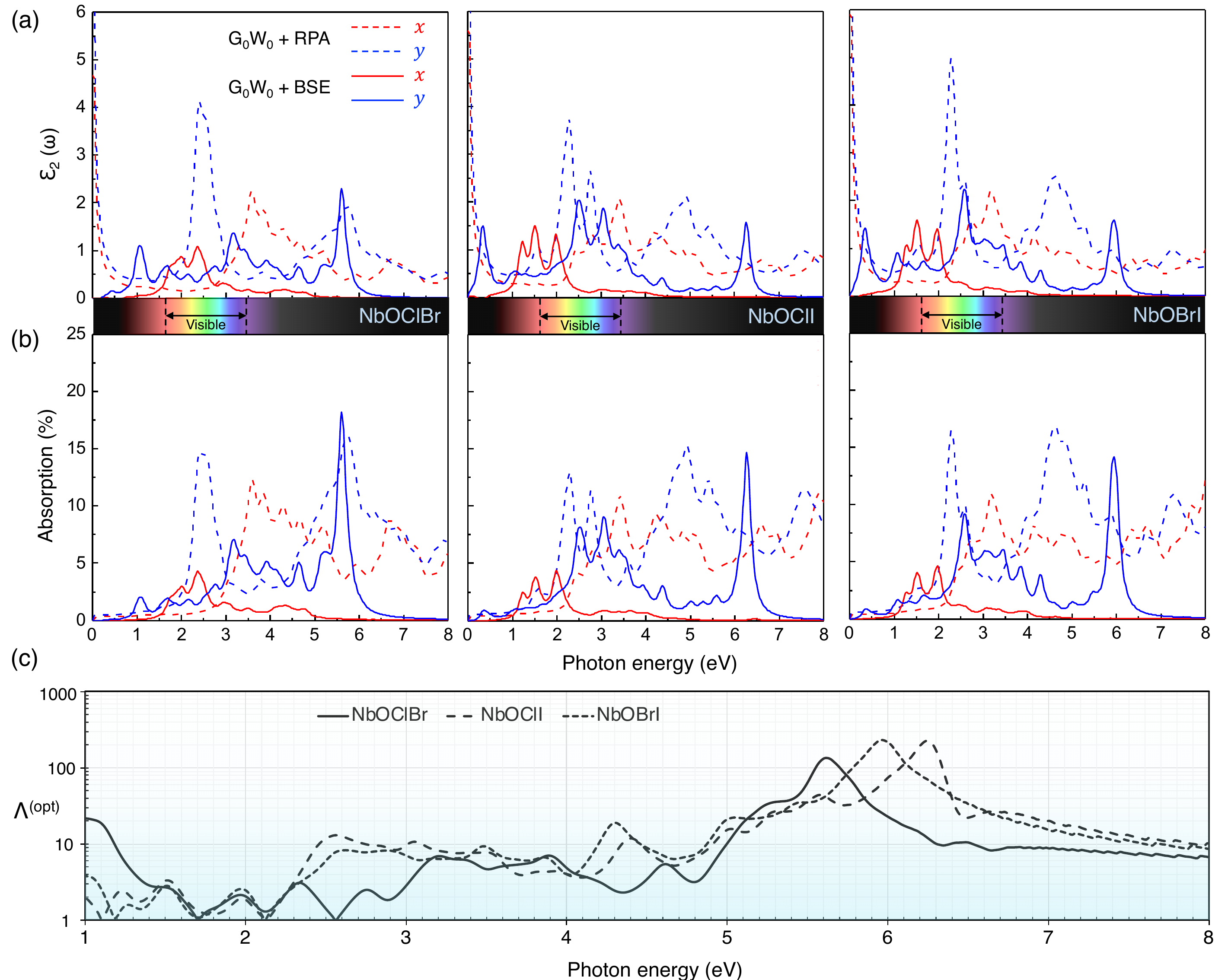}
    \caption{\textbf{Strong optical absorption and linear optical dichroism in NbO$XY$ monolayers}. (a) Imaginary part of the dielectirc function. (b) Absorption spectra. (c) Optical anisotropy ratio as a function of photon energy.}
    \label{fig6}
\end{figure*}

\subsection{Optical properties}

We now investigate the optical properties of NbO$XY$.
The imaginary part of the dielectric function $\varepsilon_2$ and the absorption coefficients are shown in Figs. 6(a) and 6(b), respectively, calculated using $G_0W_0$-RPA and $G_0W_0$-BSE methods. 
The optical properties obtained via the $G_0W_0$-RPA method, which omits the important excitonic effects, differ significantly from that of the $G_0W_0$-BSE method \cite{KILIC2022100792}. 
Especially in the visible regime, $G_0W_0$-RPA severely overestimates the optical absorption of NbO$XY$.
Since $G_0W_0$-BSE method exhibit better agreement with experiments \cite{bernardi2013extraordinary}, we focus our discussion on the $G_0W_0$-BSE results in the following.

In general, the optical absorption spectra are highly anisotropic and exhibit strong \emph{linear optical dichroism} between the $x$ and $y$ directions \cite{mhy}. 
The optical absorption in the infrared to red-visible regime for $x$-polarized light is stronger than that for $y$-polarized light. 
In contrast, the optical absorption of $y$-polarized light is stronger than that of $x$-polarized light in the blue-visible to ultraviolet regime. 
The nearly-full coverage of the infrared-visible-UV spectral regime by $x$-and $y$-directional optical responses thus suggests the potential of NbO$XY$ in solar energy conversion applications, and can complement the conversion efficiency of photocatalytic water splitting in NbOClI and NbOBrI as discussed above.

The strong linear optical dichroism of NbO$XY$ originate from their anisotropic electronic dispersion.
The band structure disperses less in the $y$ direction when compared to that in the $x$ direction, thus resulting in an overall stronger optical absorption along the $y$ direction \cite{jia2019niobium}.
We quantify the optical anisotropy as \cite{fang20212d,wang2017short},
\begin{equation}
    \Lambda^{(\text{opt})} (\hbar\omega) = \frac{ \text{max} \left[ \alpha_x(\hbar\omega), \alpha_y(\hbar\omega) \right] }{ \text{min} \left[ \alpha_x(\hbar\omega), \alpha_y(\hbar\omega) \right] },
\end{equation}
where $\alpha_{x,y}(\hbar\omega)$ is the absorption coefficient of $(x,y)$ direction, respectively, at photon energy $\hbar\omega$. 
NbO$XY$ exhibit significant $\Lambda^{(\text{opt})} \sim 10$ in the visible regime (i.e. $\hbar\omega \gtrsim 2$ eV), thus offering an avenue to straightforwardly determine the crystal direction using polarization optical microscope \cite{qiao2014few}.
In the deep UV regime, an exceedingly large $\Lambda^{(\text{opt})} > 10^2$ can be achieved.
At around $500$ nm (i.e. $\hbar\omega \approx 2.48$ eV), we obtain $\Lambda^{(\text{opt})} = 1.8, 11.3, 5.5$ for NbOClBr, NbOClI, and NbOBrI, respectively, which are comparable or larger than that of NbOI$_2$ ($\Lambda^{(\text{opt})} = 1.75$) \cite{fang20212d}, PdSe$_2$ monolayer ($\Lambda = 1.09$) \cite{lu2020layer} and GeSe ultrathin film ($\Lambda = 1.09$) \cite{wang2017short}. 
The anisotropic optical absorption of NbO$XY$ monolayers suggest their potential for polarization-sensitive photonics applications, such as linear polarizers and polarization-dependent photodetectors.

We further note that a remarkably strong optical absorption peak is present at the deep UV regime ($> 5 eV$) for NbO$XY$ monolayers. 
The peak absorption reaches 18.2 \%, 14.7 \% and 14.2\% for NbOClBr, NbOClI and NbOBrI monolayers, respectively, at the peak frequencies of 5.6 eV, 6.3 eV and 5.9 eV. 
Such absorption peaks are significantly stronger and `deeper' in the UV regime as compared to the vast majority of previously studied 2D materials \cite{2D_abs} [see Fig. 1(h)], thus suggesting the potential of NbO$XY$ for UV photonics and photodetector applications. 

\begin{figure*}
    \centering
    \includegraphics[scale=0.5058]{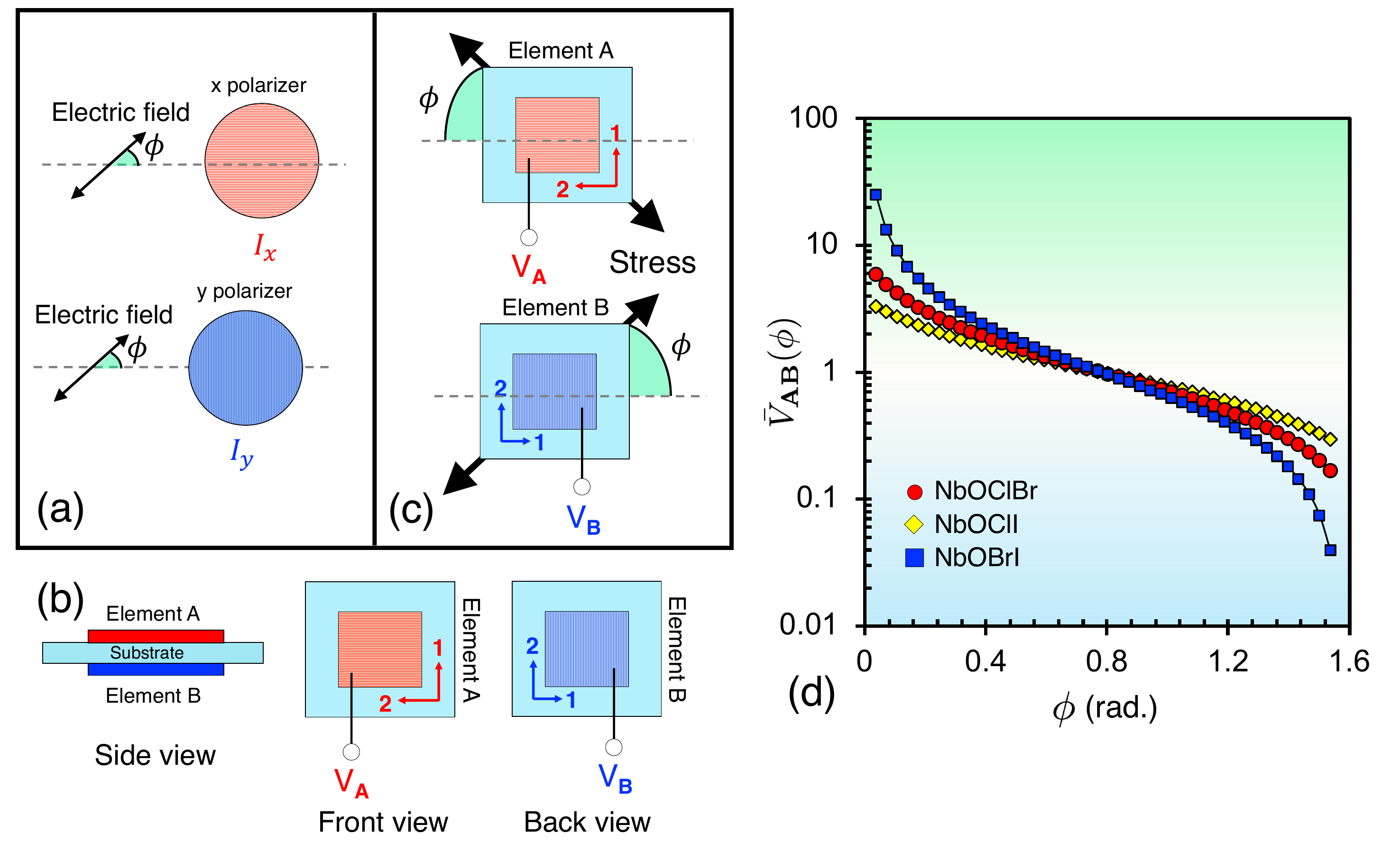}
    \caption{\textbf{Concept of directional stress sensing based on out-of-plane piezoelectricity in NbO$XY$.} (a) Schematic drawing showing the determination of electromagnetic wave polarization angle $\phi$ using two polarizer measurements. (b) Schematic drawing of the proposed directional stress sensing devices composed of two back-to-back NbO$XY$ sandwiching a flexible substrate. (c) Directional stress sensing mechanism for a mechanical stress applied at angle of $\phi$. (d) $\bar{V}_{\mathbf{AB}}(\phi)$ as a function of $\phi$ shows distinctive value for each $\phi$, thus allowing the direction of the mechanical stress to be determined. }
    \label{fig:my_label}
\end{figure*}

\subsection{Proposal of piezoelectric directional stress sensing}

Although 2D out-of-plane piezoelectric materials have been extensively studied computationally, how such properties can be harnessed for device applications remain largely unexplored thus far. Here we illustrate how the $d_{31}$ and $d_{32}$ can be harnessed for \emph{directional mechanical stress sensing}.
The operation of the proposed device (see Fig. 7 for a schematic illustration) is analogous to an electromagnetic wave polarizer. 
Consider an incoming incident linearly polarized electromagnetic wave of unknown polarization angle $\phi$ [see Fig. 7(a)]. The $\phi$ can be determined straightforwardly using two perpendicularly oriented polarizer which selectively reads out the $x$ and $y$ components of the wave intensity, i.e. $I_x$ and $I_y$, and hence $\phi = \tan^{-1} I_y/I_x$. 
Analogously, two NbO$XY$ piezoelectric elements, $\mathbf{A}$ and $\mathbf{B}$, can be stacked onto a flexible substrate with their crystal axis perpendicular to each other [see Fig. 7(b) for the side, back and front views of the proposed device]. 
Here the elements $\mathbf{A}$ and $\mathbf{B}$ serve the same role as that of the orthogonal polarizers in the electromagnetic analogy.
Consider a tensile stress applied purely along the $x$-direction, the element $\mathbf{A}$ is mechanically stressed along the `2' axis which produces larger electrical response via $d_{32}$ process, while the element $\mathbf{B}$ is mechanically stress along the `1' axis with smaller electrical response via $d_{31}$ process. 
A comparison of these signals generated from $\mathbf{A}$ and $\mathbf{B}$ thus allows the direction of the stress to be determined. 

More generally, for a mechanical stress applied along the plane of the device, $\mathbf{T} = \left( T \cos\phi, T \sin\phi \right)^{\mathcal{T}}$ where $T$ is the magnitude of the stress, $\phi$ is the angle with respect to the `2' axis of element $\mathbf{A}$ and $\mathcal{T}$ denotes transpose, the displacement field induced on element $\mathbf{A}$ is $\mathbf{D} = \mathbf{d} \mathbf{T}$ where $\mathbf{d}$ is the piezoelectric strain tensor. 
Consider only the electrical output generated on the plannar surface of element \textbf{A} due to the out-of-plane piezoelectric responses, the generated polarization charge density on the planar surface of element \textbf{A} is $\varrho_{\mathbf{A}} = d_{31} T \sin\phi + d_{32} T\cos\phi$. 
Similarly, the polarization charge density generated on the surface of element \textbf{B} is $\varrho_{\mathbf{B}} = d_{31} T \cos\phi + d_{32} T\sin\phi$.
The voltage generated on elements \textbf{A} and \textbf{B} can then be estimated based on a planar diode model, i.e. $V_{\mathbf{A},\mathbf{B}} = \varrho_{\mathbf{A}, \mathbf{B} } t / k\varepsilon_0$ where $t$ and $k$ are the thickness and the out-of-plane electrostatic dielectric constant of the 2D piezoelectric material. 
To asses the performance of the directional stress sensing, we can define the \emph{voltages response ratio} and the \emph{differential voltage response} as
\begin{subequations}
\begin{equation}
    \bar{V}_{\mathbf{AB}} (\phi) \equiv \frac{V_\mathbf{A}}{ V_{\mathbf{B}} } = \frac{ \sin\phi + \xi \cos\phi}{  \cos\phi + \xi \sin\phi},
\end{equation}
\begin{equation}
    \Delta V_{\mathbf{AB}} (\phi) \equiv  V_{\mathbf{A}} - V_{\mathbf{B}} = ( d_{32} - d_{31})  \frac{Tt}{k\varepsilon_0} (\cos\phi - \sin\phi).
\end{equation}
\end{subequations}
As $\bar{V}_{\mathbf{AB}}(\phi)$ and $\Delta V_{\mathbf{AB}} (\phi) $ are unique functions of $\phi$, measuring one of them allows the mechanical stress direction $\phi$ to be unambiguously determined [e.g. see Fig. 7(d) for the angular plot of $\bar{V}_{\mathbf{AB}}(\phi)$].
Interestingly, Eq. (13b) reveals the importance of having a sizable $d_{32}$ in achieving sensitive directional mechanical stress sensing that is otherwise not achievable in non-Janus NbOX$_2$.
Here $\Delta V_{\mathbf{AB}}$ should be, ideally, as large as possible so to achieve good signal-to-noise ratio in the measurements.
For materials with $d_{32} = 0$, $\left|\Delta V_{\mathbf{AB}} (\phi) \right| \propto d_{31}$.
In this case, a large $d_{31}$ is required to achieve large $\Delta V_{\mathbf{AB}}$.
In the contrary case of $d_{32} \neq 0$, the magnitude of the differential voltage follows $\left|\Delta V_{\mathbf{AB}} (\phi)\right| \propto \left|d_{32} - d_{31}\right|$, thus suggesting that $d_{32}$ offers an alternative route to achieve large $\left|\Delta V_{\mathbf{AB}} (\phi)\right|$ for systems in which $d_{31}$ is inherently small, such as Janus TMDC. 
To illustrate this aspect, we define the \emph{maximum differential voltage response per unit stress} rescaled by $k$ as $\Delta \tilde{\mathcal{V}}_{\mathbf{AB}}^{(\text{(max)})} \equiv k \Delta V_{\mathbf{AB}}^{\text{(max)}} / T $ where $\Delta V_{\mathbf{AB}}^{\text{(max)}}$ is the maximum of the magnitude of $\Delta V_{\textbf{AB}}$ evaluated at $\phi = 0$ or $\pi/2$. 
For NbOClBr, NbOClI and NbOBrI, $\Delta \tilde{\mathcal{V}}_{\mathbf{AB}}^{(\text{max})} = (0.10, 2.20, 1.63) \times 10^{-2}$ V/GPa, respectively.
These values are significantly larger than the $\tilde{\mathcal{V}}_{\mathbf{AB}}^{(\text{max})} = 1.15 \times 10^{-3}$ V/GPa of Janus MoSeTe monolayer (thickness $t = 3.405$ {\AA} \cite{yang2019emerging}) in which only $d_{31} = 0.030$ pm/V is present.
In general, to obtain the same $\Delta \tilde{\mathcal{V}}_{\mathbf{AB}}^{(\text{max})}$ of a piezoelectric material of thickness $t'$ that possesses only $d_{31}'$, a piezoelectric materials with simultaneous $d_{31}$ and $d_{32}$ requires only
\begin{equation}
    d_{31} =   \frac{t'}{t} \frac{ d_{31}' }{ \left| \xi - 1 \right| }
\end{equation}
where $\xi \equiv d_{32} / d_{31}$. The presence of $d_{32}$ (or equivalently $\xi$) relaxes the requirement of $d_{31}$ and offer an alternative way to achieve sensitive directional sensing of mechanical stress. 
Taking NbOClI as an example, achieving the same $\Delta \tilde{\mathcal{V}}_{\mathbf{AB}}^{(\text{max})}$ would require a hypothetical piezoelectric material of the same thickness with a much larger $d_{32} \approx 0.48$ pm/V. 
The above analysis thus highlights the important role of the out-of-plane piezoelectric response $d_{32}$ in supplementing the functionality and performance of 2D piezoelectric and piezoelectronic devices. 

\section{Conclusion}

In summary, we computationally demonstrated NbO$XY$ monolayers as a stable and mechanically flexible 2D semiconductor family with exceptional electronic, piezoelectric, photocatalytic and optical properties. 
The high carrier mobility and the band gap value lying in the visible light regime of $\sim 2$ eV suggested the strong potential of NbO$XY$ in electronic and optoelectronic device applications. 
The sizable in-plane ($d_{11}$) and two distinctive out-of-plane piezoelectric responses ($d_{31}$ and $d_{32}$) greatly enhances the design flexibility of NbO$XY$-based devices.
The strong linear optical dichroism in the visible-to-UV regime and the large optical absorption peak response in the deep UV regime revealed the capability of NbO$XY$ in photonics device applications, such as polarizer and ultraviolet light photodetector. 
Finally, the high carrier mobility, sizable built-in electric field, strong optical absorption in the visible light regime, and the appropriate band edge energies that are compatible with overall water splitting suggested the potential of NbO$XY$ in achieving high-performance solar-to-hydrogen conversion. 
Our findings unveiled NbO$XY$ as an intriguing multifunctional 2D semiconductor fmaily with strong potential in electronics, optoelectronics, UV photonics, piezoelectronics and sustainable energy applicaitons, and shall form a harbinger for the exploration of the broader 2D oxyhalides family towards next-generation advanced functional device technology. 

\section*{Author credit statement}
Tong Su: Conceptualization, Methodology, Formal analysis, Investigation, Data Curation, Writing - Review \& Editing. Ching Hua Lee: Investigation, Writing - Review \& Editing. San-Dong Guo: Writing - Review \& Editing. Guangzhao Wang: Formal analysis, Writing - Review \& Editing. Wee-Liat Ong: Investigation, Writing - Review \& Editing. Liemao Cao: Investigation, Writing - Review \& Editing. Weiwei Zhao: Supervision, Writing - Review \& Editing. Shengyuan A. Yang: Investigation, Writing - Review \& Editing. Yee Sin Ang: Supervision, Project administration, Funding acquisition, Conceptualization, Formal analysis, Writing - Original Draft, Writing - Review \& Editing, Visualization, Resources

\section*{Declaration of competing interest}
The authors declare that they have no known competing financial interests or personal relationships that could have appeared to influence the work reported in this paper.

\section*{Data availability}
Data will be made available on reasonable request.

\section*{Acknowledgement}

This work is funded by the Singapore University of Technology and Design (SUTD) Kickstarter Initiatives (SKI 2021\_01\_12) and SUTD-ZJU IDEA Visiting Professor Grant (SUTD-ZJU (VP) 202001). T. S. is supported by the China Scholarship Council (CSC). W. Z. is supported by the Natural Science Foundation of China (No. 52073075) and Shenzhen Science and Technology Program (Grant No. KQTD20170809110344233). The computational work for this article was partially performed on resources of the National Supercomputing Centre, Singapore (https://www.nscc.sg).

\end{document}